\documentclass[aps,prl,showpacs,twocolumn,superscriptaddress,preprintnumbers,amsmath,amssymb]{revtex4}

\usepackage{graphicx} 
\usepackage{dcolumn}  
\usepackage{subfigure}

\newcommand{\cc}{\ensuremath{(c\bar{c})_{\rm{res}}}}
\newcommand{\ncc}{\ensuremath{(c\bar{c})_{\rm{non-res}}}}
\newcommand{\RM}{\ensuremath{M_{\rm{recoil}}}}
\newcommand{\aprod}{\ensuremath{\alpha_{\rm{prod}}}}
\newcommand{\ahel}{\ensuremath{\alpha_{\rm{hel}}}}
\newcommand{\tprod}{\ensuremath{\theta_{\rm{prod}}}}
\newcommand{\thel}{\ensuremath{\theta_{\rm{hel}}}}

\begin{document} 

\title{Further study of double charmonium production
in $e^+e^-$ annihilation at Belle}

\author{P. Pakhlov}

\address{ITEP, B.Cheremushkinskaya, 25, Moscow, Russia\\E-mail: pakhlov@itep.ru \\
(For the Belle Collaboration)}

\begin{abstract}
\noindent
{ We report a new analysis of double
charmonium production in $e^+ e^-$ annihilation using a data sample
collected by the Belle experiment. We confirm our previous observation
of the processes $e^+ e^- \to J/\psi \eta_c (\chi_{c0}, \eta_c(2S))$
and perform an angular analysis for these processes. Processes of the
type $e^+ e^- \to \psi(2S) (c \bar{c})_{res}$ are observed for the
first time. We also observe a new charmonium state -- $X(3940)$,
produced in the process $e^+ e^- \to J/\psi X(3940)$.}
\end{abstract}

\pacs{13.66.Bc,12.38.Bx,14.40.Gx}

\maketitle
\setcounter{footnote}{0}

\noindent
The surprisingly large rate for processes of the type $e^+ e^-\to
J/\psi\,\eta_c$ and $J/\psi\,\ncc$ observed by Belle~\cite{2cc}
remains unexplained. In the Belle analysis with a data sample of
$45\,$fb$^{-1}$, the presence of the process $e^+ e^-\to
J/\psi\,\eta_c$ was inferred from the $\eta_c$ peak in the mass
spectrum of the system recoiling against the reconstructed $J/\psi$ in
inclusive $e^+ e^- \to J/\psi\,X$ events. Following the publication of
this result, the cross-section for $e^+ e^-\to J/\psi\,\eta_c$ via
$e^+ e^-$ annihilation into a single virtual photon was calculated to
be $\sim 2 \,\mathrm{fb}$~\cite{bra_excl}, which is at least an order
of magnitude smaller than the measured value. Several hypotheses have
been suggested in order to resolve this discrepancy. In particular,
the authors of Ref.~\cite{bra_psi} have proposed that processes
proceeding via two virtual photons may be important. Other
authors~\cite{gold} suggest that the final states observed by Belle
contain a charmonium state and a $M \sim 3\,\mathrm{GeV}/c^2$
glueball, which are predicted by lattice QCD. Possible glueball
contributions to the $\chi_{c0}$ signal are also discussed in
Ref.~\cite{hagi}.  In this paper we report an extended analysis of the
$e^+ e^- \to J/\psi \, \cc$ process to check the above hypotheses and
provide extra information that might be useful to resolve the puzzle.
This study is performed using a data sample of $155
\,\mathrm{fb}^{-1}$ collected around the $\Upsilon(4S)$ resonance with
the Belle detector at the KEKB asymmetric energy $e^+ e^-$ storage
rings.

\begin{figure}
\includegraphics[width=0.5\textwidth] {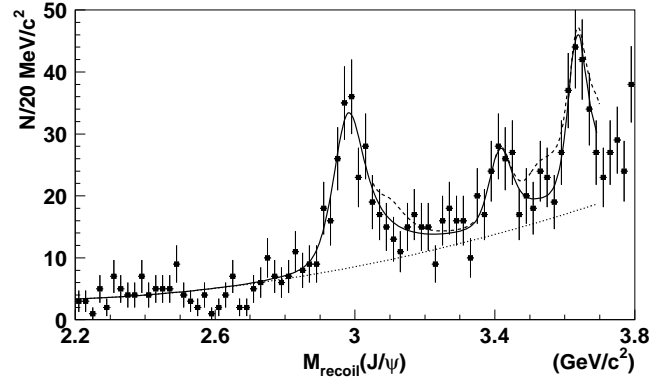}
\caption{The mass of the system recoiling against the reconstructed
$J/\psi$. The curves are described in the text.}
\label{cc7}
\end{figure}

The analysis procedure is described in detail in Refs.~\cite{2cc,prd}.
The recoil mass $\RM$ is defined as $\sqrt{(E_{\rm CM}-E_{J/\psi}^*)^2
- p_{J/\psi}^{*~2}}$, where $E_{J/\psi}^*$ and $p_{J/\psi}^*$ are the
$J/\psi$ center-of-mass (CM) energy and momentum, respectively.  The
$\RM(J/\psi)$ spectrum for the data is presented in Fig.~\ref{cc7}:
clear peaks around the nominal $\eta_c$
and $\chi_{c0}$ masses are evident; another significant peak around
$\sim 3.63\,\mathrm{GeV}/c^2$ is identified as the $\eta_c(2S)$.  The
authors of Ref.~\cite{bra_psi} estimated that the two-photon-mediated
process $e^+e^-\to J/\psi\, J/\psi$ has a significant
cross-section. To allow for a possible contribution from the exchange
of two virtual photons, we fit the spectrum in Fig.~\ref{cc7}
including all of the known narrow charmonium states below
$D\overline{D}$ threshold.  The fit results are listed in
Table~\ref{cc3t}. The yields for $\eta_c$, $\chi_{c0}$, and
$\eta_c(2S)$ have statistical significances between 3.8 and 10.7.  The
fit returns negative yields for the $J/\psi$ and $\psi(2S)$; the
$\chi_{c1}$ and $\chi_{c2}$ yields are found to be consistent with
zero. A fit with all these contributions fixed at zero is shown as a
solid line in Fig.~\ref{cc7}; the dashed line in the figure
corresponds to the case where the contributions of the $J/\psi$,
$\chi_{c1}$, $\chi_{c2}$ and $\psi(2S)$ are set at their 90$\%$
confidence level (CL) upper limit values; the dotted line is the background
function.
\begin{table}
\vbox{
\caption{Summary of the signal yields ($N$), charmonium masses ($M$)
and significances ($\sigma$).}
\label{cc3t}
\begin{center}
\begin{tabular}{l|r@{\hspace{+0.1cm}}c@{\hspace{+0.1cm}}c}
\cc & \multicolumn{1}{c}{$N$} & $M\,[\mathrm{GeV}/c^2]$& $\sigma$ \\ \hline

$  \eta_c$ & $ 235 \pm 26 $ & $ 2.972 \pm 0.007 $ & 
$  10.7  $  \\

$  J/\psi$ & $-14 \pm 20$ & fixed & ---  \\

$ \chi_{c0}$ & $89 \pm 24 $ & $3.407 \pm 0.011 $ & $3.8$  \\

$ \chi_{c1}\!+\!\chi_{c2} $ & $10 \pm 27$ & fixed & ---  \\

$ \eta_c(2S)$ & $164 \pm 30$ & $3.630 \pm 0.008 $ & $6.0$ \\

$ \psi(2S)$ & $-26 \pm 29$ & fixed & --- 
\end{tabular}
\end{center}
}
\end{table}

Given the arguments in Ref.~\cite{bra_psi}, it is important to check
for any momentum scale bias that may shift the recoil mass values and
confuse the interpretation of peaks in the \RM\ spectrum.  We use $e^+
e^-\to \psi(2S) \gamma$, $\psi(2S) \to J/\psi\,\pi^+\pi^-$ events to
calibrate and verify the recoil mass scale. From the study of the
spectrum of recoil masses squared against $\psi(2S)$ in the data, we
calculate that the $J/\psi$ recoil mass is shifted by not more than
$3\,\mathrm{MeV}/c^2$. As an additional cross-check we fully
reconstruct double charmonium events.  We find 3 pure events with
$J/\psi\,\eta_c$ combinations with energies consistent with total CM
energy.  Based on the $\eta_c$ yield in the \RM$(J/\psi)$
distribution, we expect $2.6\pm 0.8$ fully reconstructed events,
consistent with the observed signal. Thus we conclude that the peak in
\RM$(J/\psi)$ is dominated by $J/\psi\, \eta_c$ production. We also
search for fully reconstructed events with $J/\psi \, J/\psi$
combinations and find no such candidates in our data.  Based on the
calibration of the \RM$(J/\psi)$ scale, the result of the fit to the
$\RM(J/\psi)$ distribution and the full reconstruction cross-check, we
confirm our published observation of the process $e^+ e^- \to J/\psi\,
\eta_c$ and rule out the suggestion of Ref.~\cite{bra_psi} that a
significant fraction of the inferred $J/\psi \, \eta_c$ signal might
be due to $J/\psi\,J/\psi$ events.

\begin{figure}
\includegraphics[width=0.5\textwidth]{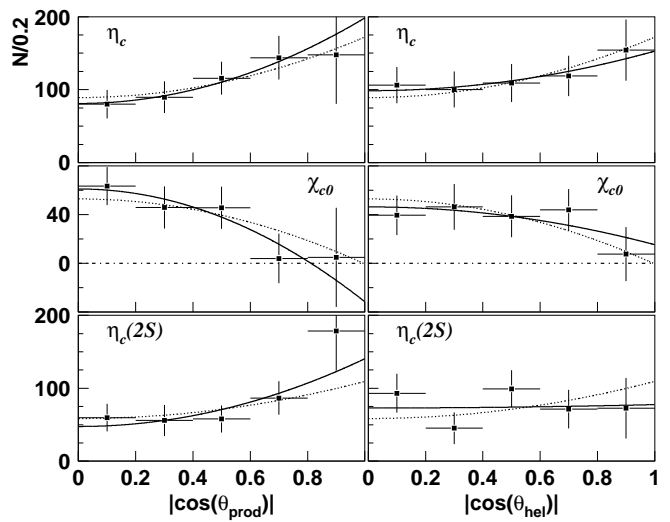}
\caption{Distributions of cosines of the production (left) and
$J/\psi$ helicity angles (right) for $e^+ e^- \to J/\psi\,\eta_c$ (top
row), $e^+ e^- \to J/\psi\,\chi_{c0}$ (middle row) $e^+ e^- \to
J/\psi\,\eta(2S)$ (bottom row). The solid lines are results of the
individual fits; the dotted lines are the simultaneous fit results.}
\label{angl}
\end{figure}

The reconstruction efficiencies for the $J/\psi\,\eta_c$,
$J/\psi\,\chi_{c0}$, and $J/\psi\,\eta(2S)$ final states strongly
depend on \tprod, the production angle of the $J/\psi$ in the CM frame
with respect to the beam axis, and the helicity angle \thel.  We
therefore perform an angular analysis for these modes before computing
cross-sections. We fit the \RM$(J/\psi)$ distributions in bins of
$\left|\cos(\tprod)\right|$ and $\left|\cos(\thel)\right|$, and
correct the yield for the reconstruction efficiencies determined
bin-by-bin from the MC.  The results are plotted in Fig.~\ref{angl},
together with fits to functions $A(1+\alpha\cos^2{\theta})$ (solid
lines). We also perform simultaneous fits to the production and
helicity angle distributions for each of the $\cc$ states, assuming
$J/\psi\, \cc$ production via a single virtual photon and angular
momentum conservation, thus setting $\aprod \equiv \ahel$. The values
of the parameter $\alpha$ are listed in Table~\ref{ang}.
\begin{table}
\vbox{
\caption{The $\alpha$ parameters obtained from fits to the production
and helicity angle distributions for $e^+ e^- \to J/\psi \, \cc$.}
\label{ang}
\begin{center}
\begin{tabular}{l@{\hspace{+0.1cm}}|r@{\hspace{+0.1cm}}r|c}
 & \multicolumn{2}{c|}{Separate} & \multicolumn{1}{c}{Simultan.} \\
 \cc     & \, \aprod\ \, & \, \ahel\ \,  & \,
$\ahel \equiv \aprod$ \,\\
\hline
$\eta_c$     & $\phantom{-}1.4^{+1.1}_{-0.8} $  & 
$\phantom{-}0.5^{+0.7}_{-0.5}$ & $\phantom{-}0.93^{+0.57}_{-0.47}$\\
$\chi_{c0}$  & $ -1.7^{+0.5}_{-0.5} $ & 
$-0.7^{+0.7}_{-0.5} $ & $ -1.01^{+0.38}_{-0.33}$\\
$\eta_c(2S)$ & $\phantom{-}1.9^{+2.0}_{-1.2}$  & 
$\phantom{-}0.3^{+1.0}_{-0.7}$ & $\phantom{-}0.87^{+0.86}_{-0.63}$ \\
\end{tabular}
\end{center}
}
\end{table}
The angular distributions for the $J/\psi\,\eta_c$ and
$J/\psi\,\eta_c(2S)$ peaks are consistent with the expectations for
production of these final states via a single virtual photon, $\aprod
= \ahel = +1$~\cite{bra_excl}.  The prediction for a spin-0 glueball
contribution ($e^+ e^- \to J/\psi\,\mathcal{G}_0$) to the
$J/\psi\,\eta_c$ peak, $\aprod = \ahel \simeq -0.87$~\cite{gold}, is
disfavored.  The process $e^+ e^- \to \gamma^\ast \to
J/\psi\,\chi_{c0}$ can proceed via both S- and D-wave amplitudes, and
predictions for the resulting angular distributions are therefore
model dependent. Our results disfavor the NRQCD expectation $\aprod =
\ahel \simeq 0.25$~\cite{bra_excl,hagi}, and are more consistent with
$S$-wave production, where $\aprod = \ahel = -1$.

To calculate the cross-sections we fix the production and helicity
angle distributions in the MC to $1+\cos^2{\theta}$ for $J/\psi\,
\eta_c(\eta_c(2S))$, and to $1-\cos^2{\theta}$ for $J/\psi\,
\chi_{c0}$.  To reduce the model dependence of our results due to the
effect of initial state radiation, whose form-factor dependence on
$Q^2$ of the virtual photon is unknown, we calculate cross-sections in
the Born approximation.  As in Ref.~\cite{2cc}, we present our result
in terms of the product of the cross-section and the branching
fraction of the recoil charmonium state into more than 2 charged
tracks: $\sigma \times \mathcal{B}_{>2}$, where $\mathcal{B}_{>2}(\cc)
\equiv \mathcal{B}(\cc \to \, >2\,\rm{charged})$. The cross-sections
are given in Table~\ref{x-sect}.

\begin{figure}
\includegraphics[width=0.5\textwidth]{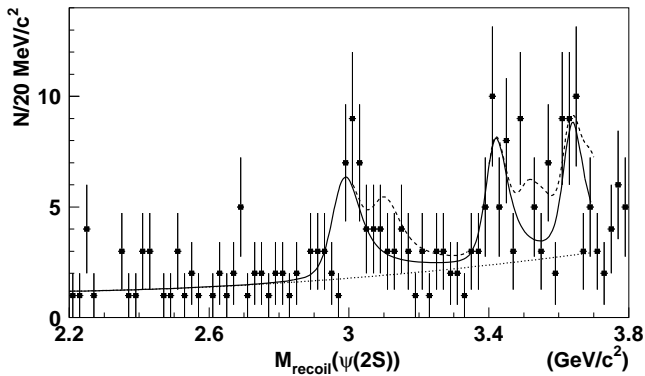}
\caption{The mass of the system recoiling against the reconstructed
$\psi(2S)$. The curves are described in the text.}
\label{cc7_2}
\end{figure}

We perform a similar study with reconstructed $\psi(2S) \to J/\psi\,
\pi^+ \pi^-$ decays to search for $e^+ e^- \to \psi(2S) \cc$ processes.
The recoil mass spectrum for the data is presented in
Fig.~\ref{cc7_2}: peaks corresponding to the $\eta_c$, $\chi_{c0}$,
and $\eta_c(2S)$ can be seen.
The fit to the $\RM(\psi(2S))$ distribution is identical to the
$\RM(J/\psi)$ fit, but due to the limited sample in this case, the
masses of the established charmonium states are fixed to their nominal
values; the $\eta_c(2S)$ mass is fixed to $3.630\,\mathrm{GeV}/c^2$ as
found from the $\RM(J/\psi)$ fit. The signal yields are listed in
Table~\ref{cc2_3t}.  Significances for the individual $\eta_c$,
$\chi_{c0}$, and $\eta_c(2S)$ peaks are in the range $3\sim4\sigma$;
the significance for $e^+e^- \to \psi(2S)\,\cc$, where $\cc$ is a sum
over $\eta_c$, $\chi_{c0}$, and $\eta_c(2S)$, is estimated to be
$5.3\,\sigma$. In Fig.~\ref{cc7_2} the result of a fit with only
$\eta_c$, $\chi_{c0}$ and $\eta_c(2S)$ contributions included is shown
as a solid line; the dashed line shows the case where the $J/\psi$,
$\chi_{c1}$, $\chi_{c2}$, and $\psi(2S)$ contributions are set at
their $90\%$ CL upper limit values; the dotted line is
the background function.
\begin{table}
\vbox{
\caption{Summary of the signal yields ($N$) and significances ($\sigma$).}
\label{cc2_3t}
\begin{center}
\begin{tabular}{l|cc}
\cc & $N$ & $\sigma$  \\ \hline

$ \, \eta_c$ & $\, 36.7 \pm 10.4 \,$ & $ \, 4.2 \, $\\

$\,  J/\psi$ & $6.9 \pm 8.9$ & ---  \\

$\, \chi_{c0}$ & $35.4 \pm 10.7 $ & $3.5$ \\

$\, \chi_{c1}\!+\!\chi_{c2} \,$ & $6.6 \pm 8.0$  &  ---  \\

$\, \eta_c(2S)$ & $36.0 \pm 11.4$ &   $3.4$ \\

$\, \psi(2S)$ & $-8.3 \pm 8.5\phantom{-}$ & --- \\ 
\end{tabular}
\end{center}
}
\end{table}
Finally, the calculated products of the Born cross-section and the
branching fraction of the recoiling charmonium state into two or more
charged tracks ($\sigma \times \mathcal{B}_{>0}$, where $
\mathcal{B}_{>0}(\cc) \equiv \mathcal{B}(\cc \to >\,0\,\rm{charged})$)
are presented in Table~\ref{x-sect}.

\begin{table}
\vbox{
\caption{Summary of the cross-sections for $e^+ e^- \to J/\psi \, \cc$
 and $e^+ e^- \to \psi(2S) \, \cc$. $\mathcal{B}_{>2(>0)}$ denotes the
 branching fraction for final states containing more than 2 (at least
 one) charged tracks. The units are fb, and the upper limits are set
 at 90\% CL. }
\label{x-sect}
\begin{center}
\begin{tabular}{l@{\hspace{+0.1cm}}rr}
\cc & $\sigma_{Born} \times \mathcal{B}_{>2}$ &
$\sigma_{Born} \times \mathcal{B}_{>0}$ \\ \hline

$\eta_c$ &  $25.6\pm 2.8\pm 3.4$ & $16.3 \pm 4.6 \pm 3.9$\\

$J/\psi$ & $<9.1$ & $<16.9$ \\

$\chi_{c0}$ & $\phantom{2}6.4\pm 1.7\pm 1.0$  &  $12.5\pm 3.8 \pm 3.1$\\

$ \chi_{c1}\!+\!\chi_{c2}\!\!\!\!$ & $<5.3$ & $<8.6$ \\

$\eta_c(2S)$ &  $16.5\pm 3.0 \pm 2.4$ & $16.0\pm 5.1 \pm 3.8$ \\

$\psi(2S)$ & $<13.3$ & $<5.2$  \\ 
\end{tabular}
\end{center}
}
\end{table}

\begin{figure}
\includegraphics[width=0.5\textwidth]{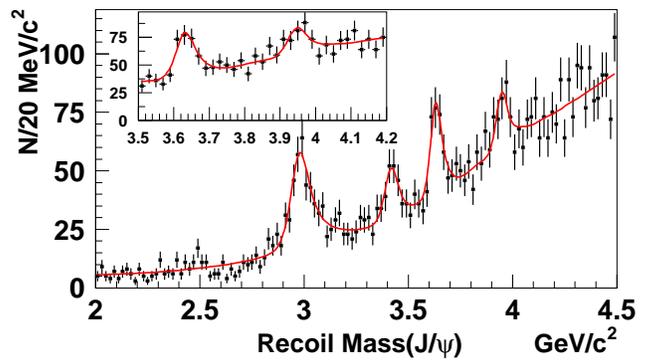}
\caption{The mass of the system recoiling against the reconstructed
$J/\psi$. The curves are described in the text.}
\label{x}
\end{figure}

Using even larger data set of $280\, \rm{fb}^{-1}$, which became
available by the summer 2004, we extend the analysis of the recoil
masses against $J/\psi$ above $D\overline{D}$ threshold. We find
another significant peak around the mass of $M \sim
3.940\,\mathrm{GeV}/c^2$ (Fig.~\ref{x}). We denote this peak as
$X(3940)$.
The fit to this spectrum includes the signal function, which is a sum
over four charmonium states: $\eta_c$, $\chi_{c0}$, $\eta_c(2S)$,
$X(3940)$, and the background function that includes also the possible
contribution from $e^+ e^- \to J/\psi D\overline{D}$ events.  The mass
of the new charmonium state found by the fit is $M_X=(3.940 \pm
0.012)\,\mathrm{GeV}/c^2$; the signal yield is $N = 149 \pm 33 $
events and the significance of the signal is $4.5\, \sigma$. The
intrinsic width of the state is consistent with zero within a large
error due to poor \RM\ resolution. We set an upper limit on $\Gamma$
to be $96\,$MeV$/c^2$ at 90\% CL. 

We also search for the decay of $X(3940)$ into $D\overline{D}$ and $D
\overline{D^*}$ final states by reconstructing one of $D$ mesons and
requiring the second $\overline{D}$ or $\overline{D^*}$ in the recoil
mass spectrum against reconstructed $J/\psi D$ combinations. We find a
significant signal of $e^+ e ^- \to J/\psi X(3940)$ process when the
event is tagged as $X(3940) \to D \overline{D^*}$, while no signal is
found for $X(3940) \to D \overline{D}$. We thus conclude that the
dominant $X(3940)$ decay mode is $D \overline{D^*}$, and this state
has a different nature from the recently found enhancement in $J/\psi
\omega$ mass distribution around the same mass~\cite{olsen}.

In summary, using a data set of $155 \,\mathrm{fb}^{-1}$ we confirm
our published observation of $e^+ e^- \to J/\psi\, \eta_c, ~ J/\psi\,
\chi_{c0}$ and $J/\psi\, \eta_c(2S)$ and find no evidence for the
process $e^+ e^- \to J/\psi\, J/\psi$. We have calculated the
cross-sections for $e^+ e^- \to J/\psi\, \eta_c$, $J/\psi\,
\chi_{c0}$, and $J/\psi\, \eta_c(2S)$ with better statistical accuracy
and reduced systematic errors and set an upper limit for $\sigma(e^+
e^- \to J/\psi\, J/\psi) \times \mathcal{B}(J/\psi \to \, >2 ~
\rm{charged})$ of $9.1\,\mathrm{fb}$ at the $90\%$ CL. Although this
limit is not inconsistent with the prediction for the $J/\psi\,
J/\psi$ rate given in Ref.~\cite{bra_psi}, the suggestion that a large
fraction of the inferred $J/\psi \, \eta_c$ signal consists of
$J/\psi\, J/\psi$ events is ruled out.  We have measured the
production and helicity angle distributions for $e^+ e^- \to
J/\psi\,\eta_c$, $J/\psi\,\chi_{c0}$, and $J/\psi\,\eta_c(2S)$; the
distributions are consistent with expectations for these states, and
disfavor a spin-0 glueball contribution to the $\eta_c$ peak.  We
observe $\psi(2S)\cc$ production for the first time, and find that the
production rates for these final states are of the same magnitude as
the corresponding rates for $J/\psi\, \cc$. Finally, using a larger
data set we observe the new narrow charmonium state at the mass
$M_X=(3.940 \pm 0.012)\,\mathrm{GeV}/c^2$.

\end{document}